\documentclass{aa}
\usepackage{graphicx}
\begin{document}

\title{The global mass function of M15}

\author{A. Pasquali\inst{1}\thanks{Present address: Institut f\"ur
Astronomie, ETH Hoenggerberg, CH-8093 Z\"urich, Switzerland} 
\and  G. De Marchi\inst{2}\thanks{ Present address: ESA, Science
Communication Office, Keplerlaan 1, 2200 AG Noordwijk,
The Netherlands} 
\and L. Pulone\inst{3}
\and M.S. Brigas\inst{1,4} 
} 

\offprints{A. Pasquali}

\institute{ESO/ST-ECF, Karl-Schwarzschild-Strasse 2, D-85748 Garching
bei M\"unchen, Germany\\
\email apasqual@eso.org 
\and
ESA, Space Telescope Operations Division, 3700 San Martin
Drive, Baltimore, MD 21218, USA\\ 
\email gdemarchi@rssd.esa.int
\and
INAF, Osservatorio Astronomico di Roma, Via di Frascati 33,
00040 Monte Porzio Catone (RM), Italy\\ 
\email pulone@coma.mporzio.astro.it
\and
Osservatorio Astronomico di Cagliari,
Strada 54, Poggio dei Pini, 09012 Capoterra, Cagliari, Italy\\ 
\email sbrigas@eso.org}

\date{Received: Submitted}

\abstract{Data obtained with the NICMOS instrument on board the Hubble
Space Telescope (HST) have been used to determine the H-band luminosity
function (LF) and mass function (MF) of three stellar fields in the
globular cluster M\,15, located $\sim 7^\prime$ from the cluster
centre. The data confirm that the cluster MF has a characteristic mass
of $\sim 0.3$\,M$_{\odot}$, as obtained by Paresce \& De Marchi (2000)
for a stellar field at $4\farcm6$ from the centre. By combining the
present data with those published by other authors for various radial
distances (near the centre, at $20^{\prime\prime}$ and at $4\farcm6$),
we have studied the radial variation of the LF due to the effects of
mass segregation and derived the global mass function (GMF) using the
Michie--King approach. The model that simultaneously best fits the LF
at various locations, the surface brightness profile and the velocity
dispersion profile suggests that the GMF should resemble a segmented
power-law with the following indices: $x \simeq 0.8$ for stars more
massive than $0.8$\,M$_\odot$, $x \simeq 0.9$ for $0.3 -
0.8$\,M$_\odot$ and $x \simeq -2.2$ at smaller masses (Salpeter's IMF
would have $x=1.35$). The best fitting model also suggests that the
cluster mass is $\sim 5.4 \times 10^5$\,M$_\odot$ and that the
mass-to-light ratio is on average  $M/L_V \simeq 2.1$, with $M/L_V
\simeq 3.7$ in the core. A large amount of mass ($\sim 44 \%$) is found
in the cluster core in the form of stellar heavy remnants, which may be
sufficient to explain the mass segregation in M15 without invoking the
presence of an intermediate-mass black hole.  

\keywords{Galaxy: globular clusters}} 

\titlerunning{The global mass function in M\,15}

\maketitle

\section{Introduction}  M\,15 (NGC\,7078) is a classical target for
studying the internal dynamics of globular clusters from the observed
surface brightness and velocity dispersion profiles for it is at a
relatively large distance from the Galactic plane ($Z_{\rm G}$ = -4.7
kpc; Harris 1996). This location, combined with an orbit of small
ellipticity ($e=0.34$; Dinescu et al. 1999), minimises the cluster
interaction with the Galaxy and hence tidal stripping and evaporation
of stars from the cluster outskirts. Ground-based observations revealed
the presence of a central cusp which has been attributed to core
collapse (Djorgovski \& King, 1986). Recent HST observations have not
been able to clarify the nature of the M\,15 cusp, in spite of their
enhanced spatial resolution. The observed stellar density profile can
in fact be reproduced by assuming the existence of either a central
black hole (Guhathakurta et al. 1996) or a compact core as the
byproduct of the cluster core collapse in the presence of diffuse dark
matter (Lauer et al. 1991). Very recently, Baumgardt et al. (2002) and
Gerssen et al. (2002) have interpreted the kinematical data obtained
with STIS and WFPC2 as due to either
strong segregation of stellar remnants (white dwarfs and neutron stars,
as already suggested by Illingworth \& King 1977)
or to the presence of a $\sim 10^3$\,M$_{\odot}$ black hole in the core
of M\,15. As before, both explanations are statistically equivalent.

\begin{table*}
\caption[]{Log of the observations}
\label{table1}
\begin{tabular}{l l c c c c c c}
\hline
       &  Dataset   &    RA (h)   &   DEC (d)  & PA (d) of the &
Filter & Number of & Total exposure\\
       &            &  J=2000   & J=2000       & detector Y axis &
       & images    &  time (s)\\
\hline
Field 1& n4k6r0hca & 21:30:17.22 & 12:15:55.4 &  0.341 & F110W & 4 & 767.835\\
       & n4k6r0hda & 21:30:17.22 & 12:15:55.4 &  0.341 & F160W & 4 & 1023.823\\
Field 2& n4k6r2hsa & 21:30:19.01 & 12:15:51.5 &  0.342 & F110W & 14 & 2687.426\\
       & n4k6r2hta & 21:30:19.01 & 12:15:51.5 &  0.342 & F160W & 14 & 3583.384\\
Field 3& n4k6u0v8a & 21:30:15.10 & 12:16:34.2 & -8.546 & F110W & 18 & 3455.262\\
       & n4k6u0v9a & 21:30:15.10 & 12:16:34.2 & -8.546 & F160W & 18 & 4607.208\\
\hline
\end{tabular}
\end{table*}

Further evidence of mass segregation in the central regions of M\,15
comes from the detection of colour gradients whereby (U-B) and (B-V)
colours get bluer towards the cluster centre (Bailyn et al. 1989;
Cederbloom et al. 1992). They have been justified in terms of either a
core concentration of blue stars due to binary-single star interactions
or the central lack of low mass main-sequence stars. In either case,
mass segregation is likely to be the driving mechanism. De Marchi \&
Paresce (1994) have resolved with HST/FOC  a large number of bright
blue stars in the core of M\,15, the majority of which can be
classified as blue stragglers. Nevertheless, the rest appear to belong
to a new, as yet unidentified class of very blue stars. Amongst several
possibilities (such as Early-Post AGB, subdwarfs and well-mixed single
stars), De Marchi \& Paresce (1994; 1996) have suggested that dynamical
interactions and close encounters could have stripped off the envelope
of red giant stars, enhancing their mass loss and heading their
evolution towards the late stage of helium white dwarf and CO white
dwarf.

Because of minimal interactions with the Galaxy, the outskirts of M\,15
have been observed to constrain the cluster initial MF (IMF) of low
mass stars. For example, De Marchi \& Paresce (1995) performed deep
HST/WFPC2 photometry of a field $4\farcm6$ NW of the centre and derived
the LF of main-sequence stars down to $M_I \simeq 10$, or two
magnitudes fainter than the LF peak. The mass distribution that Paresce
\& De Marchi (2000) have subsequently inferred from this LF shows that
the characteristic mass of M\,15 is $\sim 0.30$\,M$_{\odot}$ and that
the number of stars less massive than $0.3$\,M$_{\odot}$ quickly drops
off. This appears to be a common feature of all Galactic globular
clusters for which deep LFs are available, regardless as to their
metallicity, position in the Galaxy and dynamics (Piotto et al. 1997;
Paresce \& De Marchi 2000). This MF is believed to be representative of
the IMF of globular clusters, however the true IMF can be reliably
established only by disentangling the cluster dynamical evolution  from
the observed MF. This is best achieved when MFs are available at
several  distances from the cluster centre. For M\,15, MFs are found in
the literature near the core (De Marchi \& Paresce 1996; Sosin \& King
1997) and for the above mentioned field at $4\farcm6$ NW of the cluster
centre. Deep images taken by HST/NICMOS Camera 3 (NIC3)  during the
1998 parallel campaign have allowed us to derive the MF in three, outer
fields at 7$'$ NE of the centre. We have used this mass distribution
together with those published previously to  constrain the Global Mass
Function (GMF) of M\,15. If the  interaction  of M\,15 with the
Galactic tidal field has been as weak as recent works indicate (Gnedin
\& Ostriker 1997; Dauphole et al. 1996), then the  GMF should reflect
the IMF. 

The NIC3 observations are presented in Section\,2 and the data
reduction is described in Section\,3. The LF of the NIC3 fields in the
H band and its corresponding MF are discussed in Section\,4 and
compared with other literature measurements in Section\,5. We derive
the GMF in Section\,6 and our conclusions follow in Section\,7.

\section{Observations}

M\,15 has been observed with the NIC3 camera of HST/NICMOS  on 1998
July 7th and 18th, during the parallel observations campaign. Three
overlapping fields have been imaged at about $7^\prime$ NE from the
centre of the cluster, at a distance of 7 times the half-light radius
($1^\prime$; Trager et al. 1995). Multiple exposures have been taken of
each field through both the F110W and F160W filters, centered at
$1.1\,\mu$m and $1.6\,\mu$m, respectively. The coordinates of the
fields and the names, filters, and total exposure times of the image
data-sets are given in Table\,1. Hereafter we  refer to the F110W and
F160W bands as $J$ and $H$, respectively.

\section{Data Reduction}

The images have been reduced using the NICMOS standard calibration
pipeline: they have been first processed with CALNICA for bias
subtraction, dark-count correction and flat-fielding. Images belonging
to the same field have then been associated by means of the IRAF
routine MAKEASSOCIATION and combined with CALNICB, to remove cosmic
rays and to increase the signal-to-noise ratio.

Photometry has been performed on each of the three combined images with
the DAOPHOT package. Stars have been identified with DAOFIND, by
setting the detection threshold at $5 \sigma$ above the local
background. We have traced the radial profile of each identified object
and discarded those with full width at half maximum (FWHM) larger than
2.5 pixels, since the typical FWHM of a well defined point source in
our frames is 1.5 pixels. Moreover, we have compared images of the same
association in order to identify bad pixels not flagged by the
calibration pipeline. In this way we have selected a sample of 539
stars imaged in both $J$ and $H$ bands. Because of the highly variable
background, we decided to measure stellar count-rates in small fixed
apertures of 2 pixels in radius (equivalent to 0$\farcs4$), and the
corresponding background values in a fixed annulus with a radius of 5
pixels and a width of 2 pixels. After background subtraction and before
applying any aperture correction, we have corrected the count-rates for
the NIC3 intra-pixel sensitivity, using the equations computed by
Storrs et al. (1999; see Table\,2) in the case of out-of-focus campaign
data.

The aperture correction was determined in three steps:
\begin{itemize}

\item[i)]  First, we constructed a mean growth curve for each frame
from a sample of bright and isolated stars. The stellar fluxes were
measured in 10 apertures, with radii ranging from 1 through to 7
pixels, and the sky was taken in a fixed annulus with a radius of 7
pixels and a width of 3 pixels. After background subtraction, the
stellar count-rates obtained for the same aperture were averaged into a
mean growth curve, from which we derived the amount of energy encircled
between 2 and 5 pixels, needed to scale our count rates to an aperture
of 5 pixels.

\item[ii)] Since NIC3 was out of focus during our observations, we used
the TinyTim software (Krist \& Hook 1999) to simulate the instrumental
point spread function (PSF) with the precise optics settings
corresponding to a specific filter and observation date. We computed
two PSFs for each frame, one for our observation date (July 1998) and
one for 1998 January 15, when NIC3 was in-focus (in-focus campaigns
were carried out in January and June 1998). We calculated the encircled
energy for a 5 pixels aperture for each PSF and used the flux ratio of
in-focus and out-of-focus to correct our measured count rates.

\item[iii)]
We finally multiplied the sample count rates by the factor of $1.075$
so as to correct them to the values measured in a nominal
infinite aperture (NICMOS Photometry Cookbook, cf.
http://www.stsci.edu/hst/nicmos/performance).

\end{itemize}

The corrected count rates $c$ were then converted to magnitudes in the
VEGAMAG photometric system by means of the relation:  

\begin{equation} 
m = -2.5 log\left(\frac{cU}{Z}\right),
\end{equation}

\noindent
where $U$ is the conversion factor from flux to count rate and $Z$ is
the flux for a zero magnitude star in the VEGAMAG system, provided for
all NICMOS filters and VEGAMAG bands by the HST Data Handbook 
(http://www.stsci.edu/hst/nicmos/documents/handbooks).

\begin{table}
\caption[]{Estimated photometric errors}
\label{table2}
\begin{tabular}{l c c}
\hline
  Bin            & Error in H & Error in J\\
\hline
mag. $\leq$ 22      & 0.05  & 0.05\\
22 $<$ mag. $\leq$ 23 & 0.10  & 0.09\\
mag. $>$ 23         & 0.12  & 0.21\\
\hline
\end{tabular}
\end{table}

\subsection{Photometric uncertainty}

We estimated the internal uncertainty of our photometry by comparing
the resulting magnitudes of those stars in common to two different
fields: Fields\,1 and 2 overlap nearly over two quadrants sharing 83
stars, whilst Fields\,1 and 3 have in common only 12 stars. As a
measure of the uncertainty, we used the difference between the
magnitudes measured in each field and the weighted mean of the two
values, with the weight given by the square root of the exposure time.
The resulting errors, for Field\,2, are shown in Table\,2 for three
magnitude ranges. We associate an uncertainty of $0.05$\,mag to stars
brighter than  magnitude 22 and a photometric error $> 0.1$ to fainter
objects. As regards Field\,1, errors are usually twice as large, since
the exposure time is $\sim 3.5$ times shorter in both bands. For
Field\,3 we adopted the same uncertainty scale as for Field\,2.


\subsection{Photometric completeness}

We used the ADDSTAR routine in DAOPHOT to determine the completeness of
our photometry. We tested each frame separately in both the $J$ and $H$
band, by adding about $10\,\%$ of the total number of detected stars in
order not to increase the crowding in the images. We performed four
runs for several magnitude bins. The results are shown in Table\,3 for
all the observed fields. Since NIC3 is less sensitive in the $H$ band,
the completeness in $H$ significantly affects our subsequent data
analysis. In the case of Field\,1, for which only short exposures are
available, the completeness rapidly decreases to $\sim 50\,\%$ at $H
\simeq 23$. Fields\,2 and 3, on the other hand, reach a completeness of
$\sim 50\,\%$ and $\sim 40\,\%$, respectively, at $H \simeq 24$.

\begin{table}
\caption[]{Completeness factors}
\label{table3}
\begin{tabular}{l c c c}
\hline
 Field  &   bin  & \textit{f} (J band) & \textit{f} (H band)\\
\hline   
   1    & 15-22  &           1.        & 1.\\
   2    &        &           1.        & 1.\\ 
   3    &        &           1.        & 1.\\
\hline
   1    & 22-23   &          0.959     & 0.808\\
   2    &        &           0.949     & 0.935\\
   3    &        &           0.913     & 0.864\\
\hline
   1    & 23-23.5   &        0.807     & 0.526\\
   2    &        &           0.935     & 0.821\\
   3    &        &           0.909     & 0.680\\
\hline
   1    & 23.5-24   &        0.667     & 0.274\\
   2    &        &           0.898     & 0.545\\
   3    &        &           0.870     & 0.423\\
\hline
   1    & 24-24.5   &        0.304     & 0.126\\
   2    &        &           0.640     & 0.269\\
   3    &        &           0.433     & 0.118\\
\hline
   1    & 24.5-25   &        0.132     & 0.073\\
   2    &        &           0.488     & 0.122\\
   3    &        &           0.338     & 0.076\\
\hline
\end{tabular}
\end{table}

\section{Analysis of the NICMOS data}

\subsection{The colour--magnitude diagram}

We have corrected the observed magnitudes of our sample for reddening by
assuming E(B-V) = 0.1 from Durrell \& Harris (1993). This implies
$A_J = 0.1$ and $A_H = 0.06$. 

\begin{figure*} \centering
\includegraphics[width=15cm]{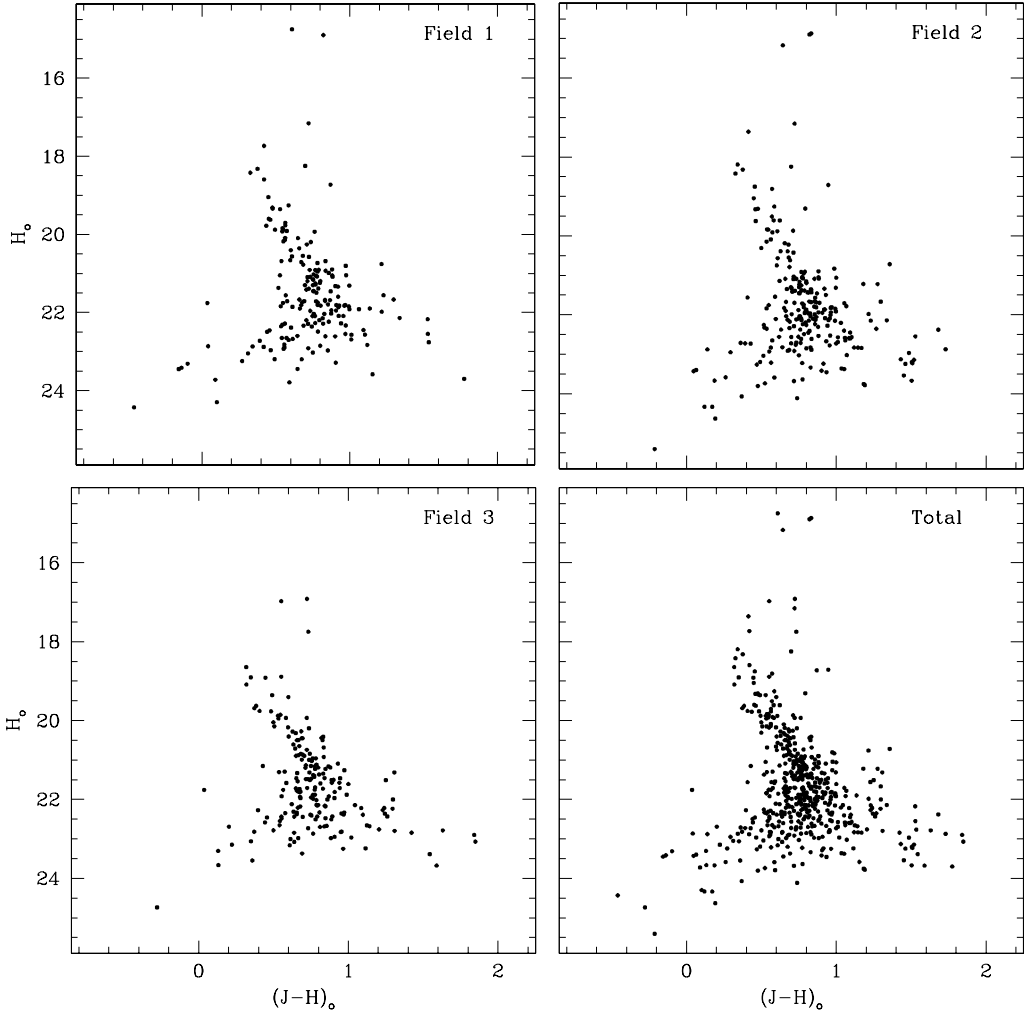} \caption{De-reddened
CMD of each observed field and all of the fields combined. Stars common
to two fields have been removed from each field but are included in the
CMD of the whole sample, with averaged magnitudes and colours.}  
\label{figure1} \end{figure*}

The de-reddened colour--magnitude diagram (CMD) is plotted for each of
the three fields and for the whole sample in Figure\,1. For the stars
in common to more than one field, we have adopted the mean magnitudes
and colours computed as above. Similar levels of photometric 
completeness are reached for all three fields at magnitudes brighter
than $H_0 \simeq 23$, so that their CMDs can be directly compared to
one another. In all the three cases the main sequence  is well defined
down to $H_0 \simeq 22$ and spreads out for $-0.4 < (J-H)_0 < 1.8$ at
fainter magnitudes due to our photometric uncertainty. The few stars
brighter than $H_0 = 18$ are probably foreground objects. Indeed,
Durrell \& Harris (1993) determined the turn-off for M\,15 at $V_0 =
19.4$ which corresponds to $H_0 = 18.4$ from the theoretical track of
Baraffe et al. (1997) at the metallicity of M\,15 ([Fe/H] = -2.15).
This implies that the stars at $H_0 \simeq 18$ are either cluster
objects evolved off the main sequence or simply foreground stars. 

The bright tip of the main sequence shows up at $H_0 \simeq 18$ for
Fields\,1 and 2 while it falls at $H_0 \simeq 19$ for Field\,3. This
apparently fainter turn-off magnitude is likely due to statistical
fluctuations in our small sample.

In order to reduce the contamination from foreground and background
stars, we applied to the CMD of each field a $2.5 \sigma$ clipping
selection around the average colour of the main sequence. The resulting
three decontaminated samples were merged onto the CMD of Figure\,2
where the stars in common are represented with averaged magnitudes and
colours. Using the predictions of Ratnatunga \& Bahcall (1985), we have
estimated the field-star contamination to be about 4 stars for each
NICMOS field in the direction of M\,15. This estimate is valid for a
limiting magnitude of $H = 23$ (i.e. $0.2$\,M$_{\odot}$ in Figure  2),
which corresponds to $V \simeq 27$ in the evolutionary tracks of
Baraffe et al. (1997), and is integrated over the whole $(B-V)$ colour
range taken into account by Ratnatunga \& Bahcall (1985). Therefore, we
do not expect  the CMD to change significantly after the
$\sigma$-clipping is applied. The photometric errors are also indicated
on the left-hand side of Figure\,2: in the range $14 \leq H_0 \leq 22$
the $(J-H)_0$ colours are known with an accuracy of $\pm 0.07$\,mag, at
$22 \leq H_0 \leq 23$ they increase to $\pm 0.13$\,mag and for $H_0 >
23$ the photometric accuracy is as poor as $\pm 0.24$\,mag.

Superposed on the observed CMD distribution is the theoretical track
obtained from the models of Baraffe et al. (1997) for a metallicity of 
$[M/H] = -2$, which closely matches the $[Fe/H] = -2.15$ value of
M\,15.  This track is scaled by the distance modulus of $15.11$ mag 
(Durrell \& Harris 1993). The stellar masses actually defining the
theoretical track are listed on the right-hand side of Figure\,2 for
decreasing $H_0$ magnitudes. The main sequence spans a mass range
between $0.8$\,M$_{\odot}$ and $0.2$\,M$_{\odot}$.

\subsection{The luminosity function}

The LF observed for the external fields of M\,15 is plotted in
Figure\,3 (solid line) corrected for incompleteness. The stars in the
CMD were grouped into magnitude bins (each $0.5$\,mag wide) between
$H_0 = 18$ and $H_0 = 24$, where completeness drops to less than
$50\,\%$. Since the completeness of the three fields is nearly the same
at $H_0 < 23$, we have assumed a mean completeness factor with which we
have corrected the star counts of bins brighter than 23. At $H_0 > 23$,
Fields\,2 and 3 are the major contributors to the observed LF, since
they are deeper. In this magnitude range we have, therefore, ignored
the  stars in Field\,1 not contained in Fields\,2 or 3 and have
rescaled the total number of objects in these two latter fields to
match the total area of the survey. Therefore we have computed a mean
completeness factor between these two fields and used the resulting
value to correct the star counts of bins fainter than $H_0 = 23$. Given
the shallower photometric depth of Field\,1, having ignored its
contribution to the LF for $H_0 > 23$ is not likely to affect the
statistical significance of our results.

\begin{figure}
\centering
\includegraphics[width=8.5cm]{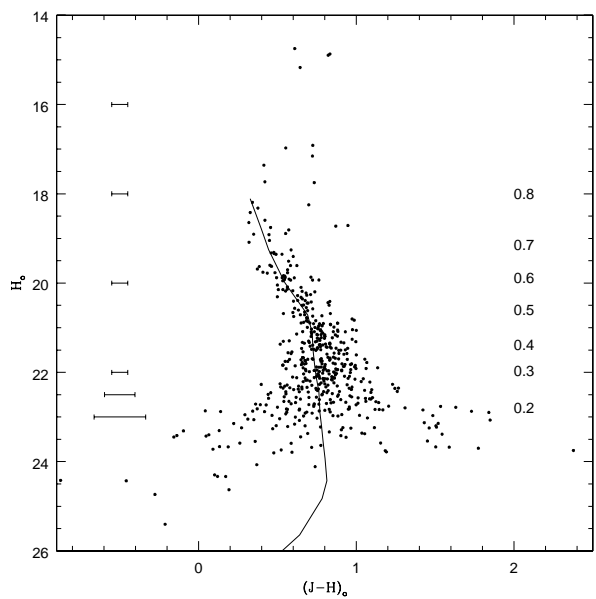}
\caption{De-reddened CMD of the whole sample on which the
theoretical track of Baraffe et al. (1997), computed for the
metallicity of M\,15 ($[Fe/H] = -2.15$), has been superposed.
Photometric errors are represented, as a function of H$_0$, at the
left-hand side of the diagram, whilst the mass range spanned by main
sequence stars is marked on the right-hand side of the plot, in unit of
solar masses.}
\label{figure2}
\end{figure}

\begin{figure}
\centering
\includegraphics[width=8cm]{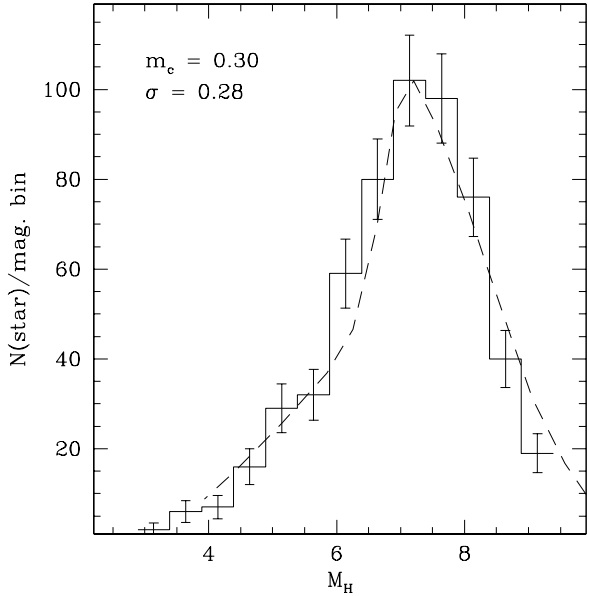}
\caption{Luminosity function (solid curve) of the whole sample, corrected for
photometric completeness. Poissonan errors have been associated to the
observed star counts. The dashed distribution is the best fitting LF,
computed under the assumption of a log-normal mass distribution with a
characteristic mass of $0.3$\,M$_{\odot}$.}
\label{figure3}
\end{figure}

The LF of Figure\,3 extends over the range $2 \leq M_H \leq 9$, peaking
at $M_H = 7.1$ (corresponding to $M_I = 8.5$; Baraffe et al. 1997) and
is characterised by a quite sharp drop to fainter magnitudes due to a
lack of progressively less massive stars. Moreover, this LF confirms
the general behaviour found by Paresce \& De Marchi (2000) for a dozen
Galactic globular clusters: their average LF, computed from stars below
1\,M$_{\odot}$ near the half-light radius, rises to a maximum value at
$M_I \simeq 8.5-9$ and then drops for fainter magnitudes (this applies
regardless of the cluster position and orbit in the Galaxy and of its
internal dynamical state).

\subsection{The mass function}

The MF of the observed fields was derived from the LF of Figure\,3.
Instead of deriving the MF by inverting the LF, we followed the
approach of Paresce \& De Marchi (2000), so as to treat separately
observational and theoretical uncertainties. We assumed a model MF of
log-normal type, i.e. one of the type:

\begin{equation} 
ln\left(\frac{dN}{dlog(m)}\right) = A -
\left[\frac{log(m/m_c)}{\sqrt{2}\sigma}\right]^2 
\end{equation} 

\noindent
with characteristic mass $m_c$ and standard deviation $\sigma$. $A$ is
a normalization constant. We then folded it through the derivative of
the mass-luminosity relationship of Baraffe et al. (1997) to obtain a
model LF, which we compared to the data until a suitable value of the
parameters was found that gives a good fit to the observations. The
best fitting LF is superposed to the observations in Figure\,3 as a
dashed distribution and is obtained with $m_c = 0.3$\,M$_{\odot}$ and
$\sigma = 0.28$.

\section{A comparison with existing surveys of M\,15}

As mentioned in the Introduction, M\,15 has been extensively studied
for it is at a relatively large distance from the Galactic plane and,
consequently, not severely affected by dynamical interactions with the
Galaxy.

\begin{figure}
\centering
\includegraphics[width=7cm]{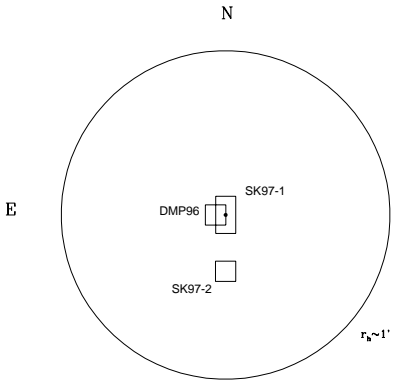}
\caption{Spatial distribution of the stellar fields observed within the
half-light radius of M\,15 ($1^\prime$;Trager et al. 1995). DMP96 is
the field studied by De Marchi \& Paresce (1996) near the core, while
SK97-1 and SK97-2 are the areas surveyed by Sosin \& King (1997).
SK97-2 is $20^{\prime\prime}$ away from the centre.}
\label{figure4}
\end{figure}

Figure\,4 spatially visualises the stellar fields observed in the core
of M\,15 by De Marchi \& Paresce (1996; DMP96) and Sosin \& King (1997;
SK97-1, SK97-2) in order to derive the central MF. The DMP96 and SK97-1
fields are at the centre, whilst the SK97-2 field is
$20^{\prime\prime}$ away from it. All fields are well within the
half-light radius of M\,15 ($r_h \simeq 1^\prime$; Trager et al. 1995).
Both studies detect a substantial amount of mass segregation, although
the LFs differ for $M_V > 4.8$. The two luminosity distributions are
plotted in Figure\,5: the LF of De Marchi \& Paresce (1996) has been
here translated from the original FOC F346M band magnitudes to standard
Johnson $V$ values using Baraffe et al.'s tracks (private
communication).

\begin{figure}
\centering
\includegraphics[width=8.5cm]{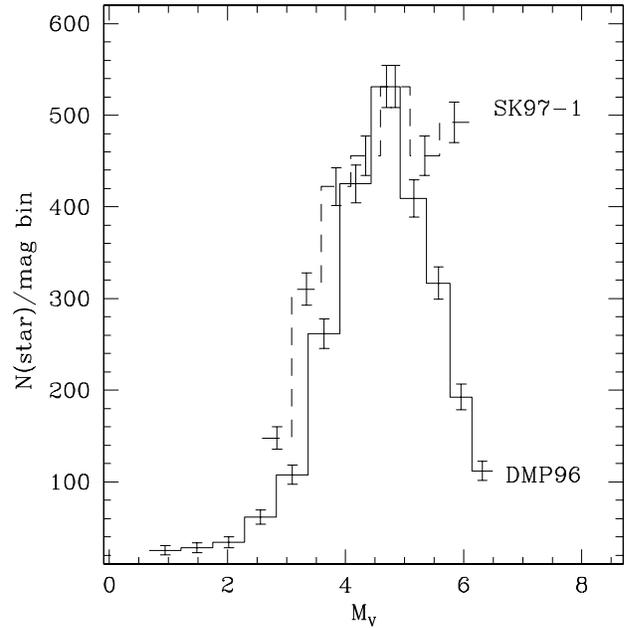}
\caption{Luminosity functions as derived by De Marchi \& Paresce (1996; solid curve) and
Sosin \& King (1997; dashed distribution) for their field positioned at
the centre of M\,15. Poisson errors have been associated with the
plotted star counts.}
\label{figure5}
\end{figure}

Both LFs have been normalised to the peak star count. An interesting
point is that at $M_V > 4.8$ the LF of De Marchi \& Paresce (1996)
dramatically drops, whilst the LF derived by Sosin \& King (1997) is
flat. We believe that this discrepancy is due to photometric
incompleteness and to the colour extrapolation from the UV to the V
band. 

Whilst mass segregation affects the innermost region of globular
clusters, dynamical interactions with the Galaxy act predominantly on
their outermost regions inducing stellar evaporation and stripping.
Thus, clusters as rich as M\,15 are likely to be dynamically
unperturbed at their half-light radius (Richer et al. 1991) and stars
at this distance can be used to constrain the IMF. Table\,4 lists the
observations available in the literature which cover the outskirts of
M\,15. Their corresponding fields are plotted in Figure\,6 together
with the cluster centre and half-light radius.

\begin{table}
\caption[]{M\,15 surveys}
\label{table4}
\begin{tabular}{l c c}
\hline
              & Distance from & Filters\\ 
              & the center    &\\
\hline
Durrell \& Harris (1993)    & $7^\prime$ NW & B,V\\
De Marchi \& Paresce (1995) & $4\farcm6$ NW & V,I\\
this work                   & $7^\prime$ NE & J,H\\
\hline
\end{tabular}
\end{table}

We extracted the LFs from these papers and homogenised them in units
of I magnitudes by using Baraffe et al.'s (1997) tracks. In Figure\,7
we over-plot the LF derived from the NIC3 data on Durrell \& Harris'
(1993; left-hand panel) and De Marchi \& Paresce's (1995; right-hand
panel) distributions. The latter were scaled to the peak star counts of
the NIC3 distribution. Unfortunately, Durrell \& Harris' (1993)
observations are not deep enough to reach the LF turn-over at $M_I
\simeq 8.5$. Nevertheless, their LF agrees well with the ascending part
of the luminosity distribution derived from the NIC3 exposures. Very
good overlap is instead achieved between the LFs of this paper and that
of De Marchi \& Paresce (1995) over the range $4 \leq M_I \leq
10$, so that both resulting MFs peak near $0.3$\,M$_{\odot}$.

\begin{figure}
\centering
\includegraphics[width=7cm]{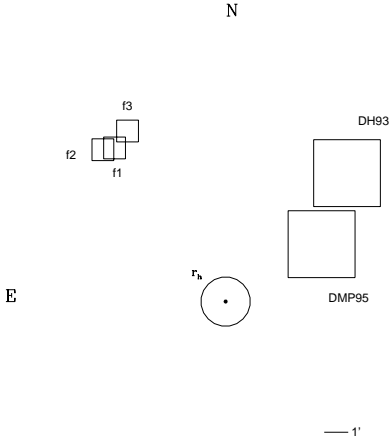}
\caption{Spatial distribution of the fields observed at distance larger
than the cluster half-light radius. DH93 labels the field imaged by
Durrell \& Harris (1993) at $7^\prime$ NW, whilst DMP95 represents the
field observed by De Marchi \& Paresce (1995) at $4\farcm6$ NW from the
centre. f1, f2 and f3 are the NIC3 fields analysed in this paper.}
\label{figure6}
\end{figure}

\begin{figure*}
\centering
\includegraphics[width=15cm]{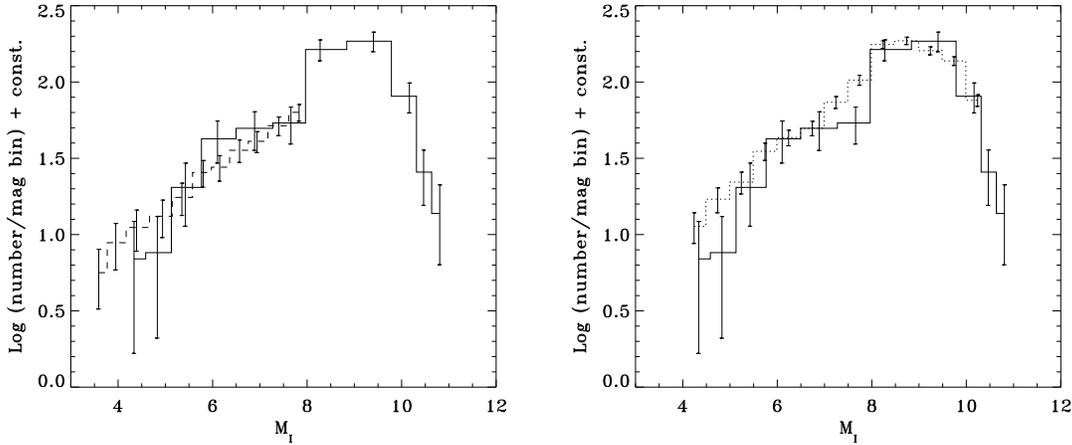}
\caption{The LFs computed by Durrell \& Harris (1993, DH93) and De
Marchi \& Paresce (1995, DMP95) are compared with the distribution
obtained from the NIC3 data. All of the luminosity distributions have
been translated into M$_I$ values and scaled to the peak star-counts of
the NIC3 luminosity function.}
\label{figure7}
\end{figure*}

\section{Dynamical structure}

Having derived the spatially resolved LF for M\,15, we can now study
its radial changes and address the issue as to whether they are
consistent with  mass segregation ensuing from two body relaxation. To
study the dynamical properties of the cluster, we employed the
multi-mass Michie--King models originally developed by Meylan (1987,
1988) and later suitably modified by Pulone et al. (1999) and De Marchi
et al. (2000) for the general case of clusters with a set of radially
varying LFs. Each model run is characterised by a MF in the form of
a power-law  $dN/d \log m \propto m^{-x}$, with a variable index
$x$, and by four structural parameters describing, respectively, the
scale radius ($ r_{\rm c}$), the scale velocity ($v_{\rm s}$), the
central value of the dimensionless gravitational potential $W_{\rm o}$,
and the anisotropy radius ($r_{\rm a}$). (After having suggested,
in Section\,3, a rather  general description of the functional form of
the MF, namely a log-normal distribution, it might seem inappropriate
to adopt a variable power-law as the basis for the MF in our dynamical
model. In fact, as we show below, the resulting  MF is
indistinguishable, for any practical purposes, from a log-normal
distribution.)

From the parameter space defined in this way, we selected those models
that simultaneously fit both the observed surface brightness profile
(SBP) and velocity dispersion profile (VDP) of the cluster as measured,
respectively, by Guhathakurta et al. (1996; for $r<100^{\prime\prime}$)
and Trager et al. (1995; for $r>100^{\prime\prime}$) and by Gebhardt et
al. (2000). However, even requiring good fits to both the SBP and VDP
can, by itself, only constrain $r_{\rm c}$, $v_{\rm s}$, $W_{\rm o}$,
and $r_{\rm a}$, whilst still allowing the MF to take on a variety of
shapes. To break this degeneracy, we further imposed the condition that
the model MF agree with the observed LF at all radial distances offered
by the data.

Since Michie--King modeling only provides a ``snapshot'' of the current
dynamical state of the cluster, it is useful to refer to the GMF, i.e.
the mass distribution of all cluster stars at present, or, in other
words, the MF that the cluster would have simply as a result of stellar
evolution (i.e. ignoring any local modifications induced by internal
dynamics and/or the interaction with the Galactic tidal field).
Clearly, in this case the IMF and GMF of main sequence (unevolved)
stars is the same. For practical purposes, the GMF has been divided
into sixteen different mass classes, covering main sequence stars,
white dwarfs and heavy remnants, precisely as described in Pulone et
al. (1999).

Our parametric modelling approach assumes energy equipartition amongst
stars of different masses. Thus, we ran a large number of trials to see
whether we could find a set of parameters for the GMF (i.e. a suitable
GMF ``shape'') such that the local MFs produced by mass segregation
would locally fit the observations. We note here that, rather than
converting the observed LFs into MFs for comparison with the
predictions of the model, we prefer to keep observational errors and
theoretical uncertainties separate. Therefore, we convert the model MFs
to LFs using for all the same M-L relation, namely that of Baraffe et
al. (1997), precisely as we did in Figure\,3. Not surprisingly, our
exercise confirms  what we had already shown in that Figure and
described above: as long as a single value of the index $x$ is used
for the GMF over the mass range $0.2 - 0.8$\,M$_\odot$, none of the
predicted local LFs can be fitted to our data. In fact, a change
of slope is needed at $m \simeq 0.3$\,M$_\odot$ so that both the rising
and dropping portions of the local LF can be simultaneously reproduced.
If we then allow the MF to take on more than one slope, the GMF that
best fits the observations is one with $x=0.9$ for stars in the range
$0.3 - 0.8$\,M$_\odot$ and $x=-2.2$ at smaller masses. The shape of
this GMF is, thus, very similar to the log-normal distribution shown in
Figure\,3. 

The set of LFs predicted by the set of Michie--King parameters
that best fit all available observations is shown in Figure\,8, where
the squares correspond to the LFs available in the literature for this
cluster at various distances from its centre. The fit to the SBP and VDP
obtained with the same set of parameter values is shown in Figures\,9
and 10 and is surprisingly good.
The values of the best fitting structural parameters
are shown in Table\,5, where they can be compared with those in the
literature. The agreement is excellent, apart from a small difference
in the value of the tidal radius which is, admittedly, not seriously
constrained by our data. We note here that we can directly compare the
observed SBP with our model since the solid line in Figure\,9
corresponds to stars of $\sim 0.8$\,M$_\odot$, namely those contributing
most of the cluster's light. As one should expect, stars in different
mass classes have different projected radial distributions.

Although stars more massive than $\sim 0.8$\,M$_\odot$ have evolved and
are no longer visible, the shape of the IMF in this mass range has
strong implications on the fraction of heavy remnants in the cluster
and, as such, on the central velocity dispersion. We find that the
best fit to the data and to the cluster's structural parameters, as
given above, requires a value of $x=0.8$ for stars in the range $100 -
0.8$\,M$_\odot$. It should be noted that the global cluster MF is thus
slightly shallower than Salpeter's IMF, which would have $x=1.35$. 
The total implied cluster mass is $\sim 5.4 \times 10^5$\,M$_\odot$ and
the mass-to-light ratio is on average $M/L \simeq 2.1$, with $M/L
\simeq 3.7$ in the core. The total cluster luminosity $L_V$ has been
estimated by integrating the best-fitting SBP (solid line in
Figure\,9), properly normalised to match the observed central surface
brightness. The best fitting models suggest that a large
fraction of mass  (of order $\sim 44 \%$) is trapped in heavy remnants,
namely stellar black holes, neutron stars and white dwarfs. However, the
presence of an intermediate mass black hole is not required.

Interestingly, the rather shallow GMF that we obtain for stars above
$0.8$\,M$_\odot$, which, in turn, results in a large fraction of heavy
remnants, is also dictated by the central enhancement seen in the SBP.
The excellent fit that our model offers to the radial surface density
of TO-mass stars (Figure\,9) would not be otherwise possible. A steeper
MF index would result in a shallower central profile which would fail
to reproduce the central density enhancement. It should also be noted
that, since a canonical King-type profile does not reproduce the SBP of
M15 (see, e.g., Trager et al. 1995), the value of $r_c$ that we obtain
does not correspond with the canonical definition of core radius,
namely that at which the  surface density falls to one-half its central
value. The latter definition only applies to the profile of the
original King model with a single mass component (see Kent \& Gunn
1982). In our case, $r_c$ simply represents a scale radius.

\begin{table}
\caption[]{Parameters of the Michie -- King models used for M\,15}
\label{table5}
\begin{tabular}{l c c c}
\hline
Parameter & Fitted & Literature & Ref.\\
           & value & value     & \\
\hline
core radius $r_{\rm c}$   & $2\farcs3$  & $2^{\prime\prime}$   & $a$\\
tidal radius $r_{\rm t}$  & $17^\prime$    & $21\farcm5$ & $b$\\
half-light radius $r_{\rm h}$ & $1\farcm3$ & $1^\prime$& $c$\\
central vel. disp. $\sigma_{\rm v}$ & 12.1 km~s$^{-1}$ & 11.7 km~s$^{-1}$
& $d$\\
\hline
\end{tabular}
\par\noindent
$a$: Guhathakurta et al. (1996)
\par\noindent
$b$: Harris (1996)
\par\noindent
$c$: Djorgovski (1993)
\par\noindent
$d$: Gebhardt (2000)
\end{table} 

\begin{figure*}
\centering
\includegraphics[width=14cm]{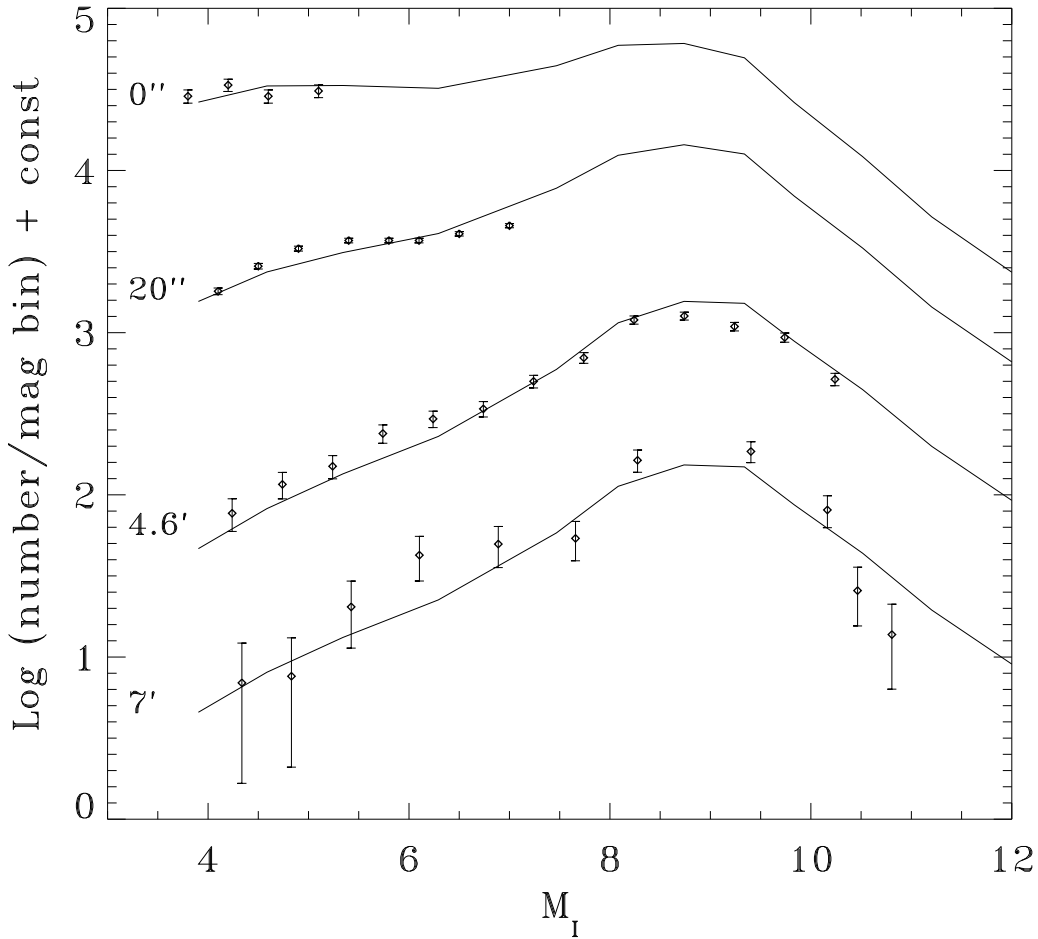}
\caption{The best fits of the Michie--King models to the LF observed
at $7^\prime$ (this work), $4\farcm6$ (De Marchi \& Paresce 1995), 
and $20^{\prime\prime}$ (Sosin \& King 1997) from the cluster centre
and in the core (De Marchi \& Paresce 1996; Sosin \& King 1997).}
\label{figure8}
\end{figure*}

\section{Other Michie--King models for M15}

Contrary to what we have concluded here, in their analysis of the
radial variation of the LF of M15, Sosin \& King (1997) concluded that
a multi-mass Michie--King model is unable to reproduced the
observations. As their Figure\,12 shows, the variation predicted by
their multi-mass model is larger than that observed when comparing the
centre of the cluster and the region at $r\simeq 5^\prime$. We have
identified three reasons that might have led Sosin \& King (1997) to
this conclusion and we discuss them here briefly. 

The first is mostly related to the approach used and, as such, should
only affect the uncertainty of the results. Sosin \& King (1997)
decided to transform the observed LFs into MFs, not necessarily using
the same M--L relation for all data, and to compare the predictions of their
multi-mass models to these MFs. The advantage of our approach, in which
the LF predicted by the model is compared with the observed LF, is that
we ensure that observational errors and theoretical uncertainty (in the
model and M--L relation) are kept separate and that only one M--L
relation is used throughout the process.

Secondly, it appears that their dynamical model is unable to reproduce
at all the observed velocity dispersion profile, as the authors
themselves point out. Conversely, ours is in excellent agreement with
the observations.
As a result, Sosin \& King (1997) predict a fraction of heavy
remnants ($\leq 1 \%$) well below the current estimates for this
cluster.

Most importantly, however, we believe that their inability to reproduce
the observed radial variation of the LF stems from the functional form
of the MF that they adopt. As we describe in Section\,4, we have made a
general assumption about the shape of the GMF, in the form of a
log-normal distribution, based on what was learnt from the observation
of a large number of halo GCs (Paresce \& De Marchi 2000). We then let
our procedure find the parameter values that simultaneously fit all
available data. Since the number of independent measurements is larger
than that of the unknowns, the procedure is bound to converge. On the
other hand, Sosin \& King (1997) adopt the MF determined by Piotto et
al. (1996) at $r\simeq 5^\prime$ as the basis for their dynamical
model, but the MF predicted by it for the cluster core fails to match
the data. The origin of the mismatch seems to lie predominantly in the
assumed shape of the MF, which is flat in the range $m \geq
0.7$\,M$_\odot$ and then sharply rises at lower masses. This results in
a MF in the central cluster regions that sharply drops in the range
$0.8$\,M$_\odot > m > 0.7$\,M$_\odot$. It appears that having adopted a
model MF with a more gentle rise from $0.8$\,M$_\odot$ all the way
through to $0.5$\,M$_\odot$ (which still fits the data at $5^\prime$
equally well) would have produced a MF in considerably better agreement
with the observations in the central cluster region.

\begin{figure*}
\centering
\includegraphics[height=13cm]{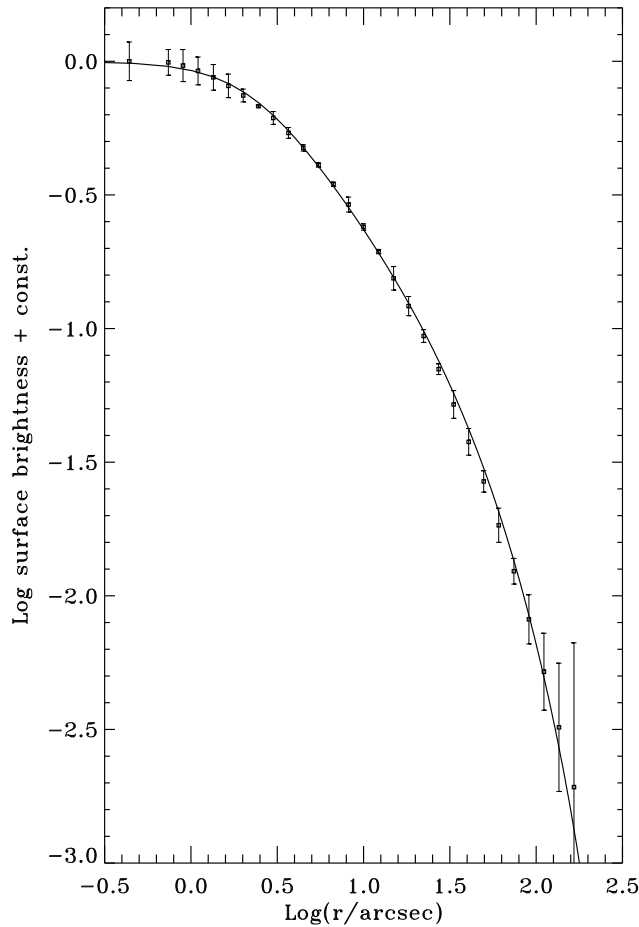}
\caption{Model fit to the surface brightness profile. The solid line
corresponds to the profile of stars of $\sim 0.8$\,M$_\odot$,
responsible for most of the light of the cluster.}
\label{figure9}
\end{figure*}

\begin{figure*}
\centering
\includegraphics[height=10cm]{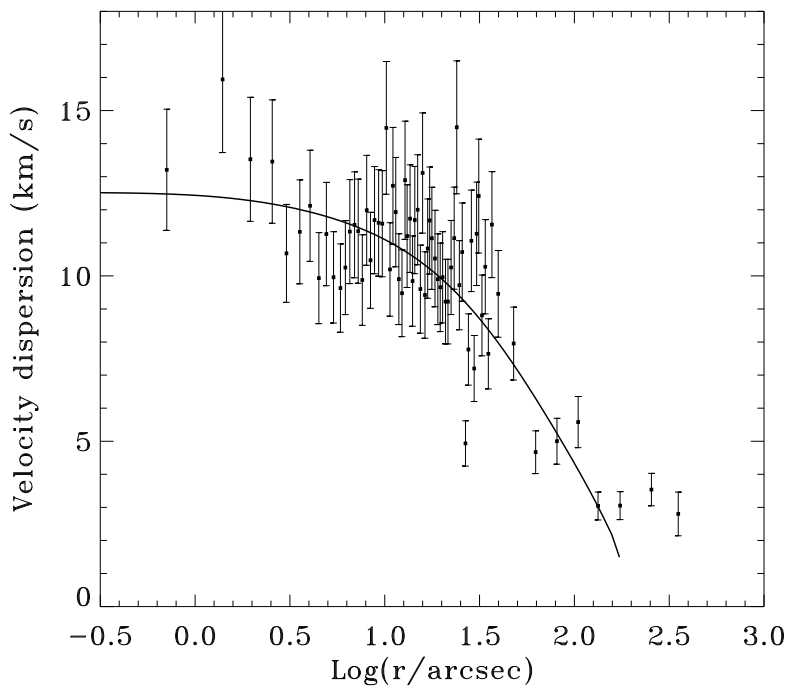}
\caption{Model fit to the velocity dispersion profile.}
\label{figure10}
\end{figure*}

\section{Conclusions}

We have analysed HST/NICMOS parallel data available for three fields in
the Galactic globular cluster M\,15 located at $7^\prime$ NE of the
cluster centre. Their total LF extends over the range $2 \leq M_H \leq
9$ and peaks at $M_H = 7.1$ (or $M_I \simeq 8.5$). It also shows a
sharp drop towards fainter magnitudes, which is a typical signature of
the lack of progressively less massive stars. We have fitted this LF
with a log-normal mass distribution and obtained a characteristic mass
of $0.3$\,M$_{\odot}$, with a standard deviation of $0.28$. These
values agree well with the characteristic mass and standard deviation
derived for a dozen Galactic globular clusters by Paresce \& De Marchi
(2000), thus supporting the ubiquity of the log-normal mass
distribution for globular clusters.

The H-band LF obtained at $7^\prime$ from the centre of M\,15 has been 
compared with the LFs derived by Durrell \& Harris (1993) and De Marchi
\& Paresce (1995) at $7^\prime$ NW and $4\farcm6$ NW of the cluster
centre, respectively. The comparison has required the translation of
all original observed magnitudes into the $I$ band. The overlap among
these three LFs is excellent, indicating that the distribution of the
stars at distances larger than the half-light radius ($\sim 1^\prime$;
Trager et al. 1995) may not be significantly perturbed in M\,15,
as expected from its Galactocentric distance ($R_G \simeq 11$\,Kpc;
Gnedin \& Ostriker 1997) and the small ellipticity of its orbit ($e =
0.32$; Dinescu et al. 1999). Indeed, calculations by Gnedin \&
Ostriker (1997) show that the time to destruction of M\,15, due to the
combined effects of internal dynamical evolution and interaction with
the Galaxy, is as large as 50\,Gyr. Therefore, if correct, this
time-scale would suggest that the outermost fields observed in M\,15
have not been significantly perturbed by tidal stripping and
evaporation and that their content should likely  represent the
{\it initial} stellar mass distribution. In other words, the high
degree of similarity between the LFs (and hence the MFs) of the fields
at $4\farcm6$ and $7^\prime$ from the centre would imply that
these are very close to be the cluster IMF and there are no significant
radial variations in the cluster IMF.

We have used the LF measured for M\,15 at $7^\prime$ (this work),
$4\farcm6$ (De Marchi \& Paresce 1995), and $20^{\prime\prime}$
(Sosin \& King 1997) from the cluster centre and in the core (De Marchi
\& Paresce 1996; Sosin \& King 1997) to study the effects of mass
segregation. We have fitted Michie--King models to the observed surface
brightness and velocity dispersion profiles in order to estimate the
cluster structural parameters and to the observed LF to constrain the
shape of the cluster GMF. The latter turns out to be characterised by
two slopes, $x=0.9$ for stars in the range $0.3 - 0.8$\,M$_\odot$ and
$x=-2.2$ at smaller masses, and is thus very close to the log-normal
distribution obtained directly from our NIC3 data near the cluster's
half-light radius.

The values of the cluster structural parameters that best fit the
observations imply a cluster total mass of $\sim 5.4 \times
10^5$\,M$_\odot$ and a mass-to-light ratio of $M/L \simeq 2.1$ on
average, with $M/L \simeq 3.7$ in the cluster core. In addition, the
best-fitting Michie--King model parameters suggest a slope of $x=0.8$
for the IMF in the range $100 - 0.8$\,M$_\odot$, which supports the
presence of a large fraction of heavy remnants ($\sim 44\,\%$) in the
core. If such a high fraction of heavy remnants is present, as
originally suggested by Illingworth \& King (1977), it would
rule out the need of an intermediate-mass black hole to explain the
mass segregation and velocity dispersions observed in the core of
M\,15.

\begin{acknowledgements}  
We are very grateful to Carlton Pryor, the referee of this paper, for
comments that have substantially strengthened the presentation of our
work. It is a pleasure to thank Isabelle Baraffe and France Allard for
providing us the theoretical tracks for the HST/FOC filters F253M and
F346M and Francesco Paresce for useful discussions. MSB acknowledges
support from the Osservatorio Astronomico di Cagliari and from the
Director General's Discretionary Fund at ESO. 
\end{acknowledgements}

\end{document}